# Microplastic and natural sediment in bed load saltation: material does not dictate the fate

Lofty, J.[1], Valero, D.[2,3], Wilson, C. A. M. E.[1], Franca, M. J. [2] and Ouro, P.[4]

[1] Cardiff University, School of Engineering, Hydro-Environmental Research Centre, Cardiff, Wales, UK

[2] Karlsruhe Institute of Technology, Karlsruhe, Germany

[3] IHE Delft, Water Resources and Ecosystems department, Delft, the Netherlands

[4] University of Manchester, School of Mechanical, Aerospace and Civil Engineering, Manchester, UK

**Abstract**

Microplastic (MP) pollution is a well document threat to our aquatic and terrestrial ecosystems, however, the mechanisms by which MPs are transported in river flows are still unknown. The transport of MPs and natural sediment in aquatic flows could be somewhat comparable, as particles are similar in size. However, it is unknown how the lower density of MPs, their shape and their different material properties impact transport dynamics. To answer this, novel laboratory experiments on bed load saltation dynamics in an open-channel flow, using high-speed camera imaging and the detection of 11,035 individual saltation events were used to identify the similarities and differences between spherical MPs and spherical natural sediments transport. The tested MPs and sediment varied in terms of size and material properties (density and elasticity). Our analysis shows that the Rouse number accurately describes saltation length, height, transport velocity and collision angles equally well for both MPs and natural sediments. Through statistical inference, the distribution functions of saltation trajectory characteristics for MPs were analogous to natural sediment with only one sediment experiment (1.4 % of cases) differing from all other plastic experiments. Similarly, only nine experiments (9.3 % of cases) showed that collision angles for MPs differed from those of natural sediment experiments. Differences observed in terms of restitution become negligible in overall transport dynamics as turbulence overcomes the kinetic energy lost at particle-bed impact, which keeps particle motion independent from impact. Overall, spherical MP particles behave similarly to spherical natural sediments in aquatic environments under the examined experimental conditions. This is significant because there is an established body of knowledge for sediment transport that can serve as a foundation for the study of MP transport.

# 1      Introduction

Plastic polymers have played an influential role in shaping human activities since the 50s - 70s of the past century (Andrady 2017; Geyer 2020) but their widespread use has also resulted in environmental contamination (Napper et al. 2020; Andrade et al. 2021; Lofty et al. 2022). Toxicologists are increasingly concerned about the potential harm caused by plastics to human health (Dick Vethaak and Legler 2021; Koelmans et al. 2022) and just recently, plastics have been found in human blood (Leslie et al. 2022), lung tissue (Jenner et al. 2022) and even the placenta (Li et al. 2018; Amereh et al. 2022; Ragusa et al. 2022). Plastics can also have detrimental effects to ecosystems on land (Browne et al. 2013; Wu et al. 2022) and water (Kirstein et al. 2016; Galloway et al. 2017; Parker et al. 2021) as particles can be readily ingested by a range of organisms (Cole et al. 2013; Wilcox et al. 2018; D'Souza et al. 2020). In addition, the capacity of plastic particles to act as vectors for pathogens, organic contaminants, and invasive species attached to their surface has triggered global concern (Gregory 2009; Viršek et al. 2017; Haegerbaeumer et al. 2019).

Plastic in the environment may be transported in air (Materić et al. 2020) or in water (Waldschläger and Schüttrumpf 2019b; Emmerik and Schwarz 2020). In the absence of vegetation (Schreyers et al. 2021; Cesarini and Scalici 2022) or in-channel structures such as a dam or a rack (Honingh et al. 2020; Meijer et al. 2021), a plastic particle in the river may be transported by the flow, however, little is known yet about plastic transport mechanisms (Waldschläger et al. 2022; Lofty et al. 2023). For instance, Waldschläger and Schüttrumpf (2019b) investigated the critical velocities for microplastic (MP) (plastics < 5 mm in size) entrainment and found significant variation up to 70 % from the classic incipient motion Shields number which was formulated for natural sediment. Other studies have focused on the settling and rising velocities of plastic particles of different shapes and sizes (Khatmullina and Isachenko 2017; Waldschläger and Schüttrumpf 2019a; Kuizenga et al. 2022), which allow the determination of drag coefficients that determine the particle–flow coupling.

Recently, it has been proposed that MP particles can be suspended in the water column in a similar way as natural sediments (Cowger et al. 2021), which was shown by Valero et al. (2022) for macroplastics (plastics > 5 mm in size) and Born et al. (2023) for nearly spherical MPs. However, there is limited understanding of the behavior of plastics in bed load transport, where particles move either by rolling/sliding or successive jumps, named saltation (Dey 2014; Ancey 2020). This is significant because the prevalence of plastics in bed load transport is comparable to that of surfaced and suspended plastics (Blondel and Buschman 2022), with up to 80% of plastic found in riverine sediments being negatively buoyant, thus potentially traveling as bed load (Hurley et al. 2018; Mani et al. 2019), and being conveyed differently to positively buoyant plastics, as suggested Born et al. (2023).

The transport of MPs and natural sediments in riverine environments could be somewhat similar, as the size of MPs are comparable to a range of natural sediments. However, it is unknown how the lower density of MPs, their shape, and their different material properties, such as elasticity, impact their bed load transport dynamics when compared to natural sediment (Waldschläger et al. 2022). In bed load, when a particle impacts with the bed with a given velocity, a portion of its momentum is lost due to restitution, which varies depending on the particle's elastic material properties (Beer et al. 2007). By losing part of the particle's velocity at impact, a different trajectory may be expected after the rebound, for instance, the particle may remain at lower depths or transition from saltation into rolling/sliding modes, potentially resulting in different concentration profiles in bed load. Therefore, a reasonable question may be posed: how will bed load transport of MPs differ when compared to natural sediment transport, for which there is an established body of knowledge available (van Rijn 1984; Garcia 2008; Ancey 2020). In order to explain the comparability between natural sediments and MPs, each of these variables need to be examined separately.

This study comparatively investigates the bed load saltation motion of spherical MPs and spherical natural sediments, which is the dominant mode of bed load transport (Wiberg and Smith 1987; Sekine and Kikkawa 1992). The material-depending variables were isolated and their influence on the transport of MPs, in terms of bed load saltation motion, was examined. Three different plastic materials in two sizes were considered, and their transport dynamics in bed load saltation were compared to experiments with amber particles, which possess similar mechanic properties to natural sediments. For direct comparability, MPs and amber were the same shape and size and experiments were undertaken using same flow conditions, methodology and analysis protocols; thus, reducing potential sources of bias. Hydrodynamic experiments were conducted in an open channel flume with controlled discharge and velocity, with discrete particle movement tracked via a high-speed camera to characterise particle behaviour. Therefore, conclusions on differences and similarities between MPs and natural sediments, uniquely from the material comparison perspective, could be drawn.

## 2 Methods

### 2.1 Experimental setup

Experiments were conducted in a 10 m long, 0.3 m wide, 0.3 m deep open channel flume with a longitudinal slope of 1/1000 (Figure 1). Rough sediment beds, made of quartz sand particles glued to plastic boards with thickness of one particle, covered the floor of the flume over the first 6 m of length, thereafter, the flume was floored with a smooth metal plate. Two sets of upstream roughness boards were tested: a first set consisted of uniformly graded sand particles with sand roughness ($k_s$) of 1.86 mm (based on median particle diameter $d50$) and geometric

---

standard deviation of the grain size ($\sigma_g = \sqrt{d_{84}/d_{16}}$) of 1.22, while a second set of boards consisted of uniformly graded sand particles with a $k_s$ of 2.76 mm and $\sigma_g$ of 1.24.

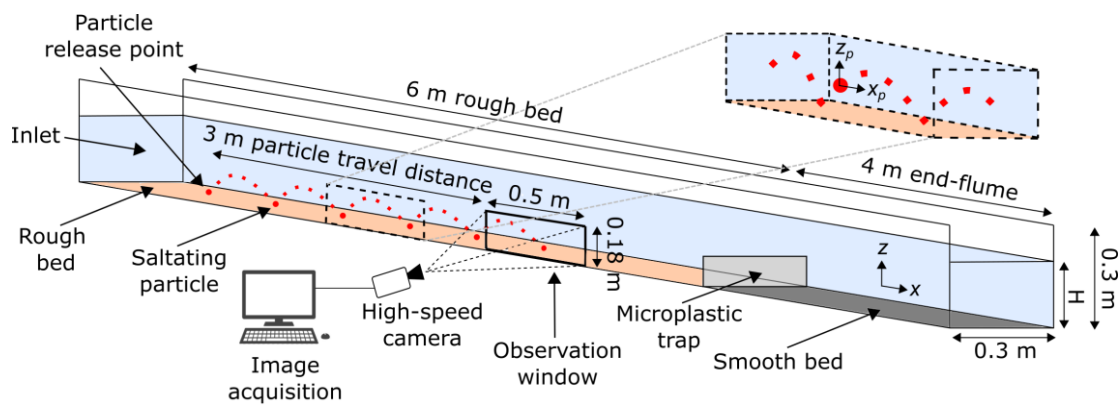

Figure 1. Experimental setup, instrumentation and particle coordinate system.

## 2.2    Flow conditions

Six steady near-uniform flow conditions for the two sets of fixed roughness beds were established in the flume. Table 1 presents their discharge ($Q$), flow depth ($H$), depth-averaged velocity ($V$), Froude number ($F = V/\sqrt{gH}$), Reynolds number ($Re = VH/\nu$), where $\nu$ is the kinematic viscosity of the fluid, shear velocity ($u_*$) based on the log-law equation and detailed velocity measurements and friction Reynolds number ($Re^* = u_* k_s/\nu$) . Flow was recirculated using a pump and measured using an electromagnetic flowmeter ($\pm$ 0.3%), while the flow depth was controlled by a weir gate. Uniformity of the flow was tested to determine whether channel flow accelerations were negligible and had little influence on the transport of particles.

Table 1. Details of the hydraulic conditions from experiments including the discharge $Q$, bed roughness height $k_s$, flow depth $H$, depth-averaged velocity $V$, Reynolds number Re, Froude number F, shear velocity $u_*$ and friction Reynolds number $Re^*$.

| $Q$ (l/s) | $k_s$ (mm) | $H$ (m) | $V$ (m/s) | F (-) | Re (-) | $u_*$ (m/s) | $Re^*$ (-) |
|---|---|---|---|---|---|---|---|
| 5 | 1.86 | 0.060 | 0.283 | 0.369 | 1698 | 0.0221 | 39.8 |
| 7.5 | 1.86 | 0.075 | 0.339 | 0.395 | 2543 | 0.0257 | 46.3 |
| 10 | 1.86 | 0.088 | 0.386 | 0.415 | 3397 | 0.0276 | 49.6 |
| 12.5 | 1.86 | 0.101 | 0.418 | 0.420 | 4222 | 0.0288 | 51.9 |
| 15 | 1.86 | 0.114 | 0.446 | 0.422 | 5084 | 0.0294 | 52.9 |
| 17.5 | 1.86 | 0.127 | 0.468 | 0.419 | 5944 | 0.0295 | 53.2 |
| 4.8 | 2.76 | 0.060 | 0.272 | 0.355 | 1632 | 0.0238 | 64.4 |
| 7.5 | 2.76 | 0.075 | 0.339 | 0.395 | 2543 | 0.0267 | 72.3 |
| 10 | 2.76 | 0.087 | 0.391 | 0.423 | 3402 | 0.0288 | 77.8 |
| 13 | 2.76 | 0.102 | 0.434 | 0.434 | 4427 | 0.0300 | 81.2 |
| 16 | 2.76 | 0.114 | 0.477 | 0.451 | 5438 | 0.0314 | 84.8 |
| 17.9 | 2.76 | 0.122 | 0.498 | 0.455 | 6076 | 0.0322 | 87.0 |

A two-dimensional particle image velocimetry (PIV) system was used for water velocity measurements in the vertical plane for each of the uniform flow conditions and bed roughnesses. The PIV system employed a high-speed Baumer VLXT-50M.I camera, able to capture images of $2448 \times 2048$ px$^2$ in size at a sampling frequency of 140 frames per second, synced with a stroboscope via a wave generator. The camera was set at 4 m downstream of the flume inlet, in the observation window (Figure 1) and captured images of $2000 \times 700$ px$^2$ ($0.50 \times 1.85$ m$^2$) in size at a sampling frequency of 120 frames per second for an interval of 30 seconds. The images were analysed using the MATLAB open-source PIV software, PIVlab (Thielicke and Sonntag 2021).

The images underwent pre-processing where background subtraction and contrast-limited adaptive histogram equalization (CLAHE) techniques were used of enhance the visibility of particles, while intensity capping and Wiener denoise filtering were used to reduce the error (Thielicke and Sonntag 2021). Image pairs were correlated by a fast Fourier transform window deformation algorithm where an interrogation window of $128 \times 128$ pixels was reduced to $32 \times 32$ pixels with a spatial overlap of 50 %. Statistical filtering was used to remove outliers departing far from the median.

The streamwise mean velocity profile was computed through spatial and time averaging. These profiles are presented for all flow conditions in Figure S1 for completeness. The law of the wall for transitionally rough beds (based on Re* in Table 1) was fitted to the velocity profiles, allowing the estimation of the shear velocity $u_*$ (Pope 2000, Eq. 7.121):

$$\frac{u}{u_*} = 5.75 \log\left(\frac{z}{k_s}\right) + B \quad (1)$$

where $u$ is the time-averaged streamwise velocity, $z$ is the distance from the bed and $B$ is a constant. The constant $B$ was determined through Fig. 7.24 of Pope (2000), which yielded values of 8.5 for channel beds that were considered hydraulically rough (Re* > 70) and 8.7 for channel beds considered in the hydraulically transitional regime (Re*= 30 − 70). The turbulent boundary layer thickness remained larger than 11.2 mm for $k_s$ = 1.86 mm, and 16.7 mm for $k_s$ = 2.76 mm, which indicates that the majority of saltation events later presented are within the boundary layer.

## 2.3 Particle properties

Three plastic materials in two different sizes (3 and 5 mm) were used in the experiments. In order of increasing density, the plastic materials considered were polyamide (PA), cellulose acetate (CA) and polyoxymethylene (POM) which are commonly observed in the riverine environment (Mani et al. 2019; Lenaker et al. 2021; Liu et al. 2021) and are denser than water (1000 kg/m$^3$), hence susceptible to be transported as bed load. **Error! Reference source not found.** presents the properties of the particles used in the experiments with details of their density ($\rho_s$), settling velocity ($W$), Young's modulus ($E$) and Poisson's ratio, defined by Cardaerlli 2008; Grote and Hefazi 2021; Tkachev et al. 2021.Click or tap here to enter text.

Table 2. Properties of the MP and amber particles used in the experiments including their diameter $d$, density $\rho_s$, setting velocity $W$, Young's modulus $E$ and Poisson's ratio defined by Cardaerlli 2008; Grote and Hefazi 2021; Tkachev et al. 2021

| Particle | $d$ (mm) | $\rho_s$ ($\pm$ std) (kg/m$^3$) | $W$ ($\pm$ std) (m/s) | $E$ (GPa) | Poisson's ratio |
|----------|----------|-------------------------------|------------------------|-----------|-----------------|
| PA | 3 | $1104.0 \pm 5.9$ | $0.0811 \pm 0.0030$ | 2.0 | 0.38 |
| | 5 | $1107.1 \pm 5.2$ | $0.1195 \pm 0.0033$ | 2.0 | 0.38 |
| CA | 3 | $1262.0 \pm 15.3$ | $0.1267 \pm 0.0043$ | 3.3 | $0.39 - 0.44$ |
| | 5 | $1262.4 \pm 7.5$ | $0.1813 \pm 0.0045$ | 3.3 | $0.39 - 0.44$ |
| POM | 3 | $1346.5 \pm 14.1$ | $0.1456 \pm 0.0039$ | 3.0 | $0.35 - 0.37$ |
| | 5 | $1361.7 \pm 7.5$ | $0.2136 \pm 0.0038$ | 3.0 | $0.35 - 0.37$ |
| Amber | 5 | $1041.3 \pm 8.3$ | $0.0717 \pm 0.0025$ | 70 | 0.30 |

For comparison, amber particles were also used in the experiments, which have been considered as a proxy for low density natural sediments in riverine transport experiments (Shields 1936; Rouse 1939) (Table 2). Table 3 presents the mechanical properties of various natural sediment materials and amber particles. It can be seen that the material properties of amber, such as the Young's modulus ($E$), which is a key parameter in particle restitution (Melo et al. 2021), is larger than plastics (Table 2), but are analogous to common natural sediments formed from rocks (Table 3).

Table 3. Material properties of amber compared to other common natural sediments (Ji et al. 2002; Cardaerlli 2008).

| Sediment type | $\rho_s$ (kg/m³) | $E$ (GPa) | Poisson's ratio | Mohs hardness |
|---|---|---|---|---|
| Amber | 1050–1090 | 70 | 0.30 | 2 - 2.5 |
| Quartz | 2640–2730 | 56–79 | 0.10 – 0.22 | 7 |
| Calcite | 2715–2940 | 72 - 88 | 0.32 | 3 |
| Dolomite | 2760–2840 | 70–91 | 0.10 – 0.35 | 3 - 4 |

Despite MPs being observed in many different shapes in the riverine environment (Hurley et al. 2018; Corcoran et al. 2020; Woodward et al. 2021), all particles used in experiments were spherical in shape to isolate the impact that the material-dependent variables have on the transport dynamics of plastic. Particles used in experiments are shown in Figure 2 against the two rough beds used.

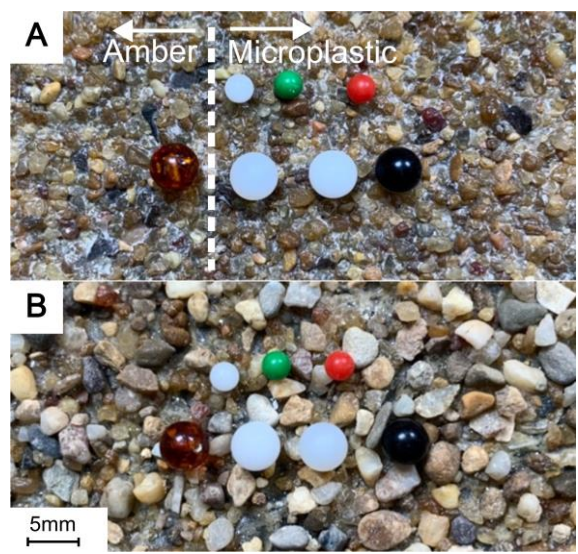

Figure 2. Photographs of the 5 mm and 3 mm MP and the 5 mm amber particles used in experiments on roughened beds with $k_s$ of A) 1.86 mm and B) 2.76 mm.

The density ($\rho_s$) of the MPs and amber were determined by using the immersion method according to DIN 53479 standards (ISO 2019). The settling velocity ($W$) was estimated for all particles, following a similar method to previous studies (Khatmullina and Isachenko 2017; Waldschläger and Schüttrumpf 2019a; Jalón-Rojas et al. 2022). A clear cylindrical plexiglass settling column of 1.5 m in height and 0.15 m in diameter was used. The column was filled with distilled water in order to minimise the interaction between the falling particles and other suspended materials in the water. MPs and amber were released by tweezers at the centre of the settling column below the surface of the water to reduce the water surface tension affecting the particles settling velocity. The high-speed camera was used to capture the falling particles at 90 frames per second over a vertical distance of 0.6 m, excluding the upper and lower 0.45 m of the cylinder height to allow the particle to accelerate to terminal velocity. The settling velocity of each MP and amber particle was then calculated as the mean velocity of the falling particle. Each test was repeated five times for each particle.

Given the particle settling velocity ($W$) and the shear velocity ($u_*$) for each flow condition, the Rouse number $P$ can be calculated as (Rouse 1939):

$$P = \frac{W}{\beta \kappa u_*} \qquad (2)$$

where $\kappa = 0.41$ is the von Kármán constant and $\beta$ is a parameter that adjusts the assumption of parabolic eddy diffusive which is assumed to be equal to unity. $P$ determines the shape of the Rouse profile, which is a theoretical concentration profile of particles in turbulent flows and is also used to determine the mode at which particles are transported by the flow. A value of $P >$ 2.5 usually indicates a particle is transported as bed load, while a value of $P$ between 0.8 - 2.5 indicates that a particle is in suspended mode of transport (Dey 2014; Cowger et al. 2021). In this study, Rouse numbers ranged between 5 to 24, which suggests that all particles should predominately be in bed load transport for all flow conditions analysed. For completeness, Shields numbers ranged between 0.027 – 0.512 and Reynolds particle numbers ranged between 66.3 - 161.1 for all flow conditions and particles used.

## 2.4 Experimental procedure

Figure 1 shows the experimental setup. Between 15 to 20 particles of the same diameter and polymer (**Error! Reference source not found.**), for every combination of bed roughness and flow condition (Table 1), were manually released in the upstream region at bed height at the centreline of the flume, one at a time. The particles travel for three meters before moving into the observation window, where the high-speed camera recorded at 90 frames per second captured the motion of the particles with a field of view of $0.5 \times 0.18$ m$^2$. The release point was deemed far enough upstream so that the particle movement captured by the camera was unaffected by the initial conditions or any disturbance developed at the release point, and the particles achieved a steady bed load motion before passing through the observation window. An example video of a 5 mm PA particle moving over a 1.86 mm roughness bed is provided within the supplementary material. At the downstream end of the experimental area, MPs were collected by a sediment trap composed of a mesh sheet (Figure 1). 11,035 individual saltation events were observed throughout all experiments.

## 2.5    Video post-processing and saltation characterisation

Video recordings were analysed using open-source software in Fiji (ImageJ2) (Gulyás et al. 2016), where the coordinates ($x_p$, $z_p$) of the centroid of the particles moving through the observation window were extracted at each frame of the recorded videos. The coordinates were then used to calculate the mode of particle transport of the particles: rolling/sliding, saltation and suspension. For the saltation events, particle trajectory characteristics and particle–bed collision characteristics were determined. Examples of a 5 mm PA particle trajectory on a 2.76 mm roughness bed, under three different flow condition are shown in Figure 3 and the different transport modes (rolling/sliding, saltating or suspended transport) are highlighted.

The different modes of particle transport were determined by an algorithm coded in R statistical software (Lofty 2023) implementing the following algorithm. Looking at a particle, the trajectory from impact to impact with the bed is considered an individual event. Saltation events are events in which the centroid of the particle exceeded a height of $k_s/d$ away from the average height of the sediment bed. If the particle did not reach such heights during the event, then it is classified as rolling/sliding, and involves the particle moving majorly in contact with the bed. Conversely, suspension events were determined as those in which the particle is reverted upwards during the falling limb of the trajectory, i.e., the particle moves upwards again in the middle of the falling trajectory and before the collision with the bed (Figure 3C, green trajectory), which can only be driven by turbulent forces that keep the particle in suspension (Abbott et al. 1977).

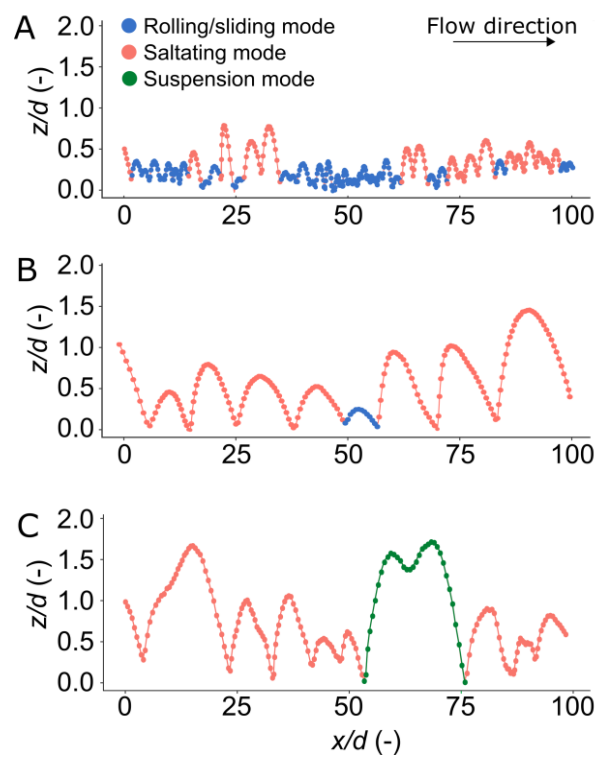

Figure 3. Trajectories of a 5 mm PA particle travelling over a roughened bed with a $k_s$ of 2.76 mm at increasing shear velocity A) $u_* = 0.0238$ m/s ($P = 12.5$), B) $u_* = 0.0288$ m/s ($P = 10.4$) and C) $u_* = 0.0322$ m/s ($P = 9.27$). Rolling/sliding, saltation and suspension events are highlighted in different colours.

To characterise the trajectories of the saltation events in each recording, the saltation length $(L_p)$ was calculated as the distance between two successive saltation collisions with the bed, the saltation height $(H_p)$ was calculated as the maximum height of the saltation event relative to the height of the bed, and the saltation transport velocity $(U_p)$ was calculated as the saltation length divided by the saltation event duration (Figure 4A). Both $L_p$ and $H_p$ were made dimensionless by the particle diameter $d$, while $U_p$ was made dimensionless by the shear velocity $u^*$.

To characterise the collision dynamics, the average inwards $(\alpha_{in})$ and outwards $(\alpha_{out})$ collision angles (*prior* and *posterior*), relative to the bed horizontal, were calculated using the $x_p$, $z_p$ coordinates immediately before and after the particle impact (Figure 4B). Similarly, the streamwise inwards $(u_p|_{in})$ and outwards $(u_p|_{out})$ velocity and the vertical inwards $(w_p|_{in})$ and outwards $(w_p|_{out})$ velocities were also calculated. The inwards $(v_p|_{in})$ and outwards $(v_p|_{out})$ collision velocity magnitude was calculated as:

$$v_p|_{in} = \sqrt{u_p|_{in}^2 + w_p|_{in}^2} \quad (3)$$

$$v_p|_{out} = \sqrt{u_p|_{out}^2 + w_p|_{out}^2} \quad (4)$$

Where $u_p$ and $w_p$ are streamwise and vertical velocities of the particles, calculated through central differences.

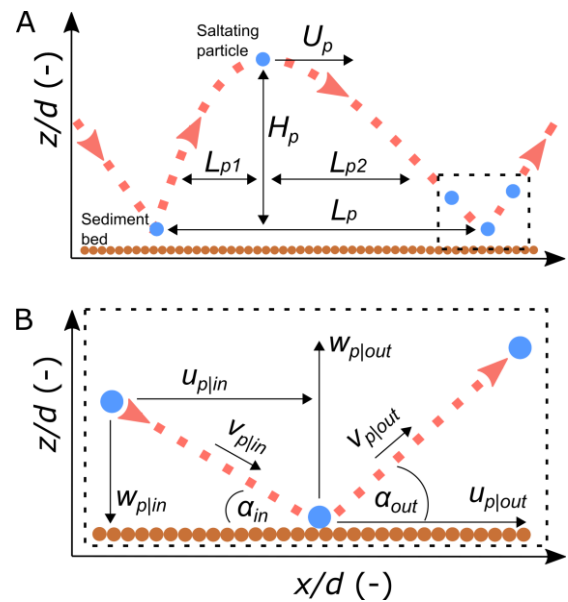

Figure 4. A) Definitions of particle trajectory characteristics and B) Definitions of particle – bed collision characteristics.

## 3    Results and discussion

### 3.1    Bed load mode of transport

A total of 1,665 individual MP and amber particle runs were recorded, with multiple modes of bed load (rolling/sliding and saltation) and suspended load transport observed. Figure 5 shows the percentage of time that particles spent either in saltation or in rolling/sliding mode of transport. All particles were mobile for all experimental conditions, which meant no particles were in repose, while suspension events remained below 5% for all particles. Given the low correlation between mode of bed load transport and the Rouse number ($R^2 = 0.223$, Figure S2), a dimensional analysis approach was undertaken to evaluate the relationship between the different physical particle properties ($d$, $\rho_s$, $W$), fluid motion ($u_*$) and bed properties ($k_s$) and the percentage of the MP and amber particles transported as bed load. Figure 5 shows that a dimensionless parameter with the configuration of a modified Rouse number, $k_s\, W/d\, u_*$ which described more consistently the relative frequency of saltation or rolling/sliding modes of transport ($R^2 = 0.817$),

The parameter, $k_s\, W/d\, u_*$ encloses two ratios of variables that define the mechanics of bed load transport of plastics: the first, which refers to the Rouse number, $(W/u_*)$ accounts for the relative influence of the particle's settling velocity $W$ to that of turbulence, described by $u_*$; the second, accounts for how the bed roughness impacts bed load transport dynamics and it is defined by the relative roughness to the size of the particle $(k_s/d)$, similar to conclusions about hiding/exposure effects made by Waldschläger and Schüttrumpf (2019b) for the initiation of motion for MPs.

The improved fit of this parameter suggests that the mode of transport is not only determined by flow and turbulence (Rouse-based considerations) but also by bed roughness and MP particle size. Hence, particles are more likely to saltate as the ratio between the roughness and MPs diameter decreases or as the strength of turbulence acting upon the particle increases. The collapse of transport mode in Figure 5 for all particles indicates that material properties did not significantly influence this parameter.

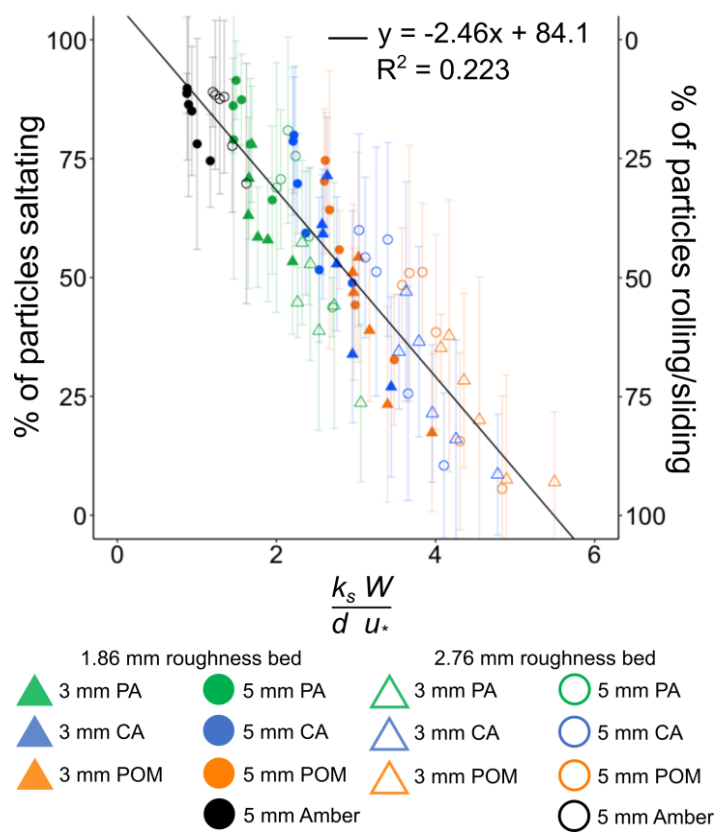

Figure 5. Percentage of particles in saltation or rolling/sliding mode of transport against a modified Rouse number $k_s W/du_*$, with the solid line indicating a linear fitting. No particles were in repose and less than 5% of all particles were in suspension mode of transport, thus omitted from this analysis.

## 3.2 Particle trajectory characteristics

Figure 6A-C shows the main descriptors of the particle trajectories (saltation length $L_p$, saltation height $H_p$ and saltation transport velocity $U_p$), which are computed as an average for each material for a certain flow condition and bed roughness. Lower Rouse numbers, indicating higher relative turbulent forces, yielded higher values of $L_p$, $H_p$ and $U_p$, suggesting a stronger particle–flow coupling. For the same flow condition, particles with a lower settling velocity (lighter and/or smaller particles) can reach higher regions in the water column with higher

velocity, which will accelerate them further; on the other hand, denser and/or larger particles tend to remain closer to the bed. Negligible differences can be seen for values of $L_p$, $H_p$ and $U_p$ between the two roughness beds, thereby indicating that the $k_s/d$ effect that was not relevant in the trajectory descriptors, as it was in the determination of the permanence of the particles in rolling/sliding or saltating modes as suggested by Figure 5A.

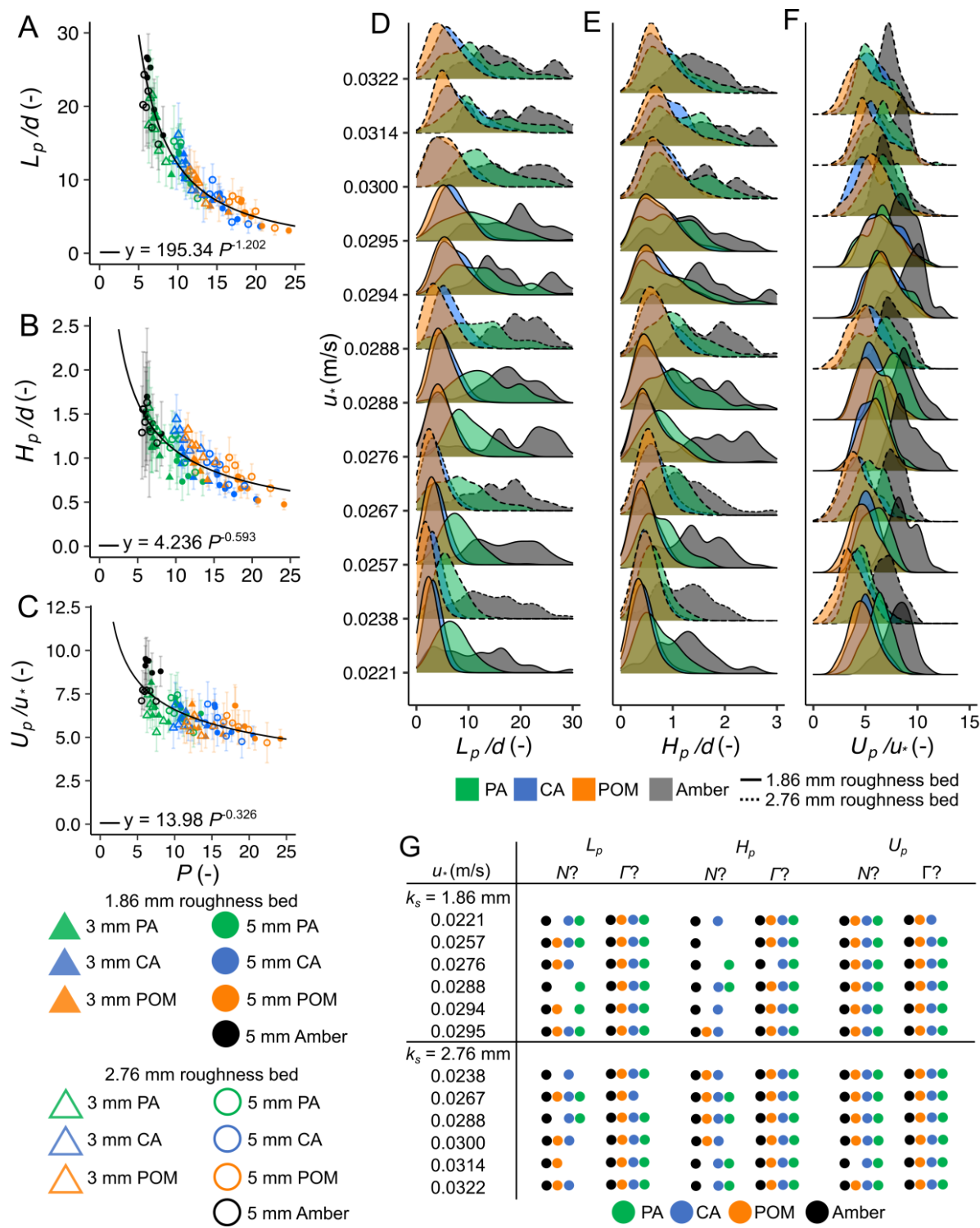

Figure 6. Average saltation trajectory characteristics, in terms of A) saltation length $L_p$, B) saltation height $H_p$, and C) saltation velocity $U_p$ plotted against the Rouse number $P$. Error bars represent the standard deviation for each material and flow condition. Kernel density plots displaying distribution of 5 mm particles for D) $L_p$, E) $H_p$, and F) $U_p$ for each flow condition $u_*$ and bed roughness $k_s$ using a kernel density bandwidth following suggestions of Silverman (1986). G) Summary of the K-S test indicating where distributions of trajectory characteristics are consistent with a Gaussian ($N$) and/or Gamma ($\Gamma$) functions i.e., p > 0.05.

Figure 6 D – F shows the frequency distribution of $L_p$, $H_p$ and $U_p$, considering all individual events for the 5 mm particles of all materials. Frequency distribution plots for the 3 mm MP particles are shown in Figure S3, for completeness. For each flow condition, the trajectory characteristics are represented through kernel density estimates, which are a nonparametric smoothing alternative to the fitting of a parametric probability density function (Wilks 2006). It is observed that with decreasing Rouse number, distributions show larger dispersion (also observable in Figure 6A-C), regardless of the material, likely due to the relatively stronger diffusive effect of turbulence upon the particles' trajectory.

Conversing theories on sediment trajectory probability density functions have been previously proposed; $L_p$ and $H_p$ have both been suggested to fit a Gaussian (Hu and Hui 1996b) and Gamma (Lee et al. 2000; Lee et al. 2010; Roseberry et al. 2012) distribution, while $U_p$ has been proposed to fit an exponential (Fathel et al. 2015; Shim and Duan 2019), which is also positively skewed, and Gaussian distributions (Hu and Hui 1996b; Lee et al. 2000; Lee et al. 2010). In this study, two functions were considered to describe the frequency distributions of

---

$L_p$, $H_p$ and $U_p$ across particle materials: Gaussian ($N$), as data may fall symmetrically around the mean, and Gamma ($\Gamma$), to take into consideration the potential skewness, or long tails, that data may show. This may be expected since particle trajectory characteristics should be physically bounded to zero; e.g., no negative saltation jump lengths or heights should be expected.

A one–sample Kolmogorov–Smirnov (K-S) test was performed to identify whether the sampled data is drawn from either a Gaussian or a Gamma function (Wilks 2006). For that purpose, a Gaussian and a Gamma function are fitted to the data via maximum likelihood estimation. Provided that the probability density functions are identified for MP and amber particles' saltation characteristics, the following question can be addressed: are MPs' trajectories different from those of natural sediments (using amber particles as a proxy)? Figure 6G compiles the results of the K-S tests for each trajectory characteristic, flow condition, roughness configuration and particle material. Each point in Figure 6G indicates if the data distribution is consistent with a Gaussian and/or Gamma function (p > 0.05).

Figure 6G can be seen as a similarity matrix in which the points in the same cell indicate that similar probability density functions may be expected for the different particles under the same flow conditions and bed roughness. For instance, considering particle velocity $U_p$, for the lower $k_s$ bed and a shear velocity of 0.0257 m/s, Gaussian and Gamma distributions can explain the data of all types of particles analogously; however, looking at the distribution of saltation height $H_p$, we see that only the Gamma function explains this data for all materials, while the Gaussian distribution only explains the amber particles. Altogether, it can be observed that distribution functions of the characteristics from amber experiments were analogous to plastic materials for 85 % of all cases, and only one case (1.4 %) is different from all the parallel plastic experiments.

From visual inspection of Figure 6, no apparent difference in behaviour can be observed between MPs and amber particle's trajectory characteristics from an averaged (Figure A-C) or frequency (Figure 6D-F) point of view. Therefore, a simple model based on $P$ should be able to capture saltation dynamics, regardless of the particle material. Following careful consideration of four regression models (linear, power law, exponential and logarithmic functions), fitted to the data via least squares error minimisation for expected values of $L_p$, $H_p$ and $U_p$, it is concluded that a power law function minimised relative bias (Bennett et al. 2013), the most accurate across all materials (Table S1). For completeness, the power law functions are included for Figure 6A-B.

### 3.3 Saltation trajectory shape

The shape of the saltation trajectory for particles in bed load is asymmetrical, with the falling limb being larger than the rising limb (Hu and Hui 1996a; Lee et al. 2000). Results suggest that $L_{p2} \approx 1.5 L_{p1}$ and that the ratio of $L_{p2}/L_{p1}$ is independent of the Rouse number, particle material and bed roughness, indicating that the shape of saltation trajectory is alike between MPs and amber particles. The maximum trajectory height ($H_p$) is roughly 40 % of the total trajectory length, which is consistent with previous studies for natural sediment (Hu and Hui 1996a; Lee et al. 2000).

---

The shape of the particle saltation trajectories can also be evaluated by identifying the relationship between the average $L_p$ and $H_p$ as shown in Figure 7A. $H_p$ grew linearly with $L_p$, with both increasing with decreasing Rouse number. This occurred for both MPs and amber particles. For visualization purposes, Figure 7B illustrates the trajectory ranges observed for 5 mm MPs and amber particles trajectories, accounting for the average minimum, maximum and values for $L_p$ and $H_p$. It is seen that the amber particles saltation is longer and higher than the MP particles, which is explained by the larger $P$ values, mainly caused by the smaller settling velocity of the amber particles.

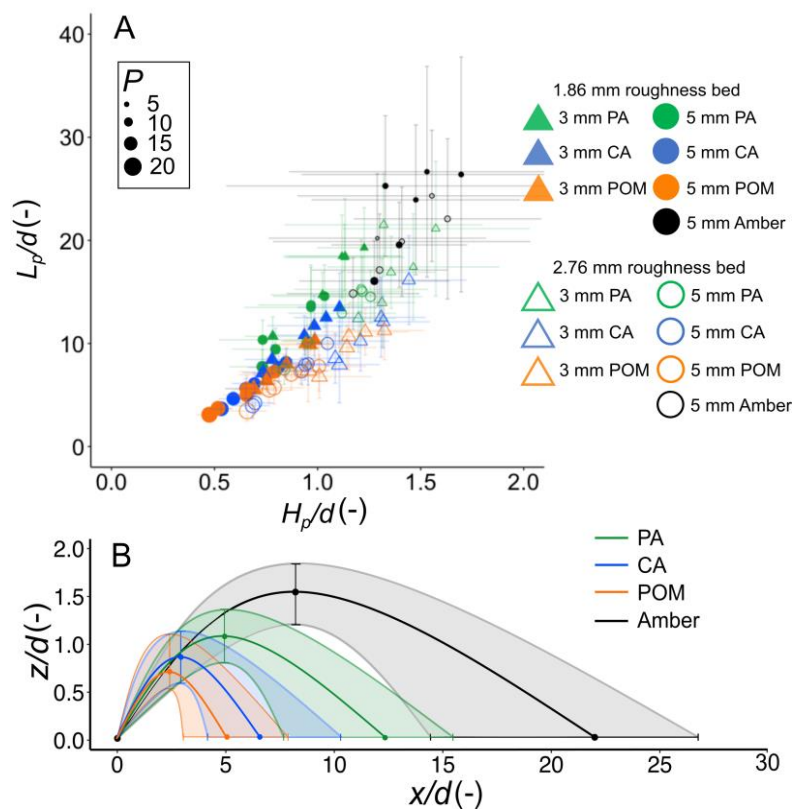

Figure 7. A) The average saltation length $L_p$ plotted against saltation height $H_p$ with point size indicating the Rouse number. Error bars represent the standard deviation. C) The average minimum, maximum and mean experimental trajectory range for 5 mm particles.

## 3.4 Particle collision characteristics

### 3.4.1 Collision angles

Particle collision characteristics are key features of particle saltation as they provide information on the energy loss during impact and rebound dynamics (Zeeshan Ali and Dey 2019; Pähtz et al. 2020). Figure 8A shows the average inwards collision angle ($\alpha_{in}$) against the outwards angle ($\alpha_{out}$) for the MP and amber particles, as defined in Figure 4. Results indicate that $\alpha_{in}$ ranged from 4.8 – 15.5° whilst values of $\alpha_{out}$ ranged between 12.2 – 32.6° (min – max,

---

for average angles). On average, $\alpha_{out}$ is larger than $\alpha_{in}$ for all particles, flow conditions and bed roughnesses, suggesting that particles are directed upwards after collision.

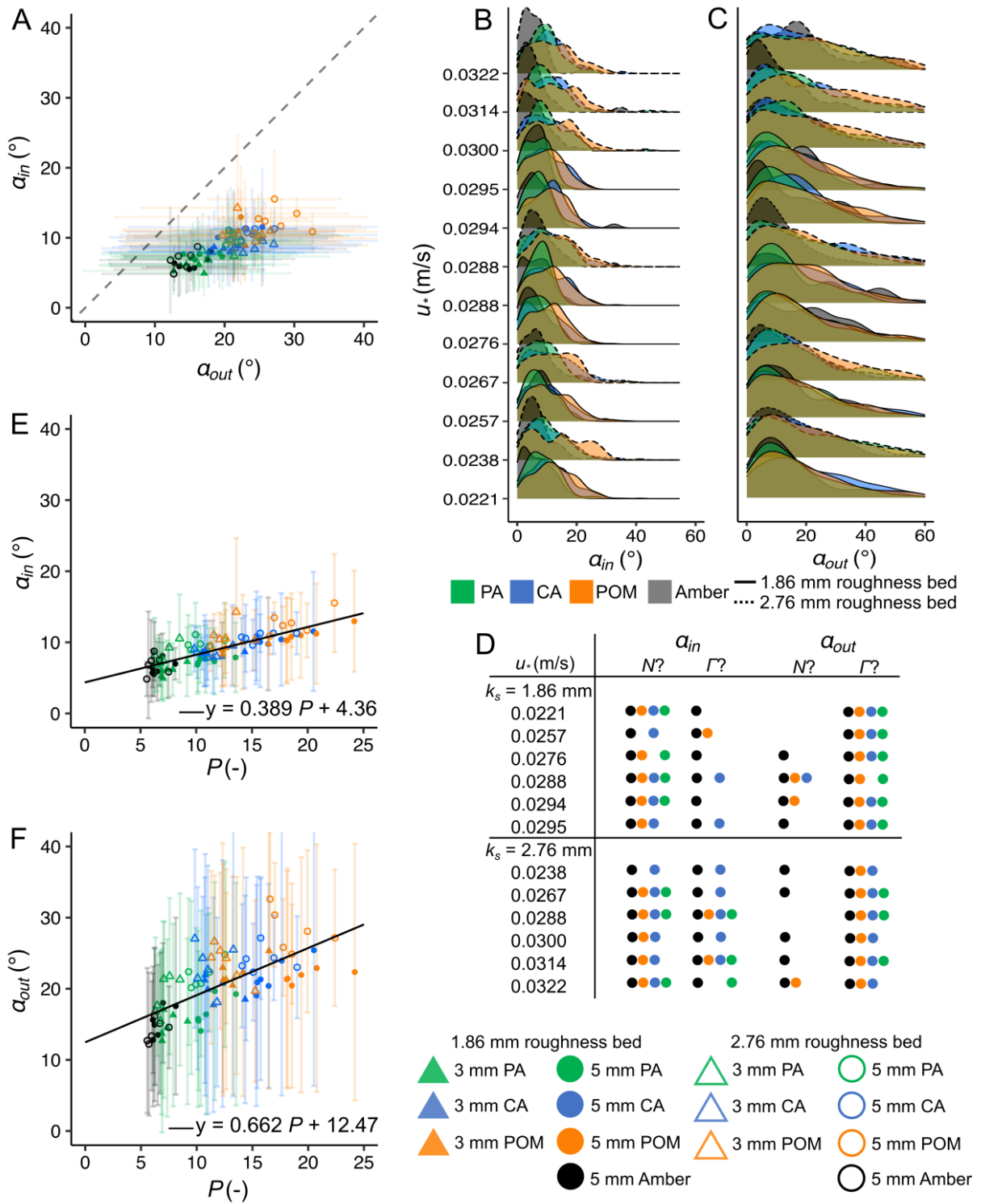

Figure 8. A) Inwards collision angle $\alpha_{in}$ plotted against the outwards angle $\alpha_{out}$ and respective kernel density plots for 5 mm particles for B) $\alpha_{in}$ and C) $\alpha_{out}$ for each flow condition $u_*$ and bed roughness $k_s$, using a kernel density bandwidth following suggestions of Silverman (1986). D) Summary of the K-S test indicating where distributions of $\alpha_{in}$ and $\alpha_{out}$ are consistent with a Gaussian ($N$) and/or Gamma ($\Gamma$) functions i.e., p > 0.05. E) The inwards collision angle $\alpha_{in}$ as a function of the Rouse number $P$ a with fitted linear function. F) The outwards collision angle $\alpha_{out}$ as a function of the Rouse number $P$ with fitted linear function.

Figures 8B-C show the frequency distribution of $\alpha_{in}$ and $\alpha_{out}$ angles for 5 mm particles represented through kernel density estimates, for each flow condition. The distributions show that, $\alpha_{out}$ has larger dispersion compared to $\alpha_{in}$, likely to be a result of particle impact with a heterogeneous roughened bed, which leads to a wider range of outward angle possibilities. Conversely, $\alpha_{in}$ is mostly a consequence of the shape of the particle trajectory that was found to be relatively uniform across flow conditions (Section 3.3). This difference in distribution between $\alpha_{in}$ and $\alpha_{out}$ angles is similar to results found by previous experimental research for natural sediments (Lee et al. 2000).

A similar analysis to Section 3.2 was undertaken to identify whether sampled $\alpha_{in}$ and $\alpha_{out}$ data is drawn from either a Gaussian or a Gamma function using K-S tests, and if there are differences between probabilistic descriptors of plastic and natural sediments. Figure 8D compiles the results of the K-S tests for $\alpha_{in}$ and $\alpha_{out}$ for each flow condition and bed roughness, showing where data is consistent with a Gaussian and/or Gamma distribution (p > 0.05). The results show that distribution functions of the collision angle from amber experiments were

---

analogous to the plastic materials for 57 % of all cases, with only nine cases (9.3 %) differing from all comparable plastic experiments.

At lower Rouse numbers, the particle – flow coupling is stronger and a particle reaches higher water levels in the water column, with larger flow velocities, thus is carried by the flow for a longer distance and this elongates the particle trajectory, causing a flatter $\alpha_{in}$ and $\alpha_{out}$ angles. Based on the observations in Figure 8E and F, $\alpha_{in}$ and $\alpha_{out}$ appear to be well described by the Rouse number. To quantify their relationship, a linear model is provided that is fitted to the data using least squares error minimisation and keeps relative bias and uncertainty (Bennett et al. 2013) similar across materials (Table S2) and is shown in Figure 8E-F for completeness.

### 3.4.2  Collision velocity

Figure 9 shows the average inwards and outwards collision velocities in the streamwise ($u_p|_{in}$, $u_p|_{out}$, Figure 9A) and vertical directions ($w_p|_{in}$, $w_p|_{out}$, Figure 9B), as well as the velocity magnitude ($v_p|_{in}$, $v_p|_{out}$, Figure 9C). Results show that in the streamwise direction, $u_p|_{in}$ is larger than $u_p|_{out}$ for all particles, while in the vertical direction, $w_p|_{out}$ is larger than $w_p|_{in}$. These results indicate that particles are ejected from collision with a transfer of momentum into the vertical component. During this impact, a part of this energy is dissipated to the bed.

It should be noted that values of $w_p|_{in}$ and $w_p|_{out}$ are a magnitude smaller than values of $u_p|_{in}$ and $u_p|_{out}$, highlighting that the streamwise component dominates the particle's transport and particle–bed collisions. Thus, there is a limited contribution of the vertical velocity component in the calculation of the of $v_p|_{in}$ and $v_p|_{out}$ at collision with the bed. As a result, $v_p|_{in}$ is greater than $v_p|_{out}$, confirming that overall particles move slower after impact with the bed as kinetic energy is lost during collision.

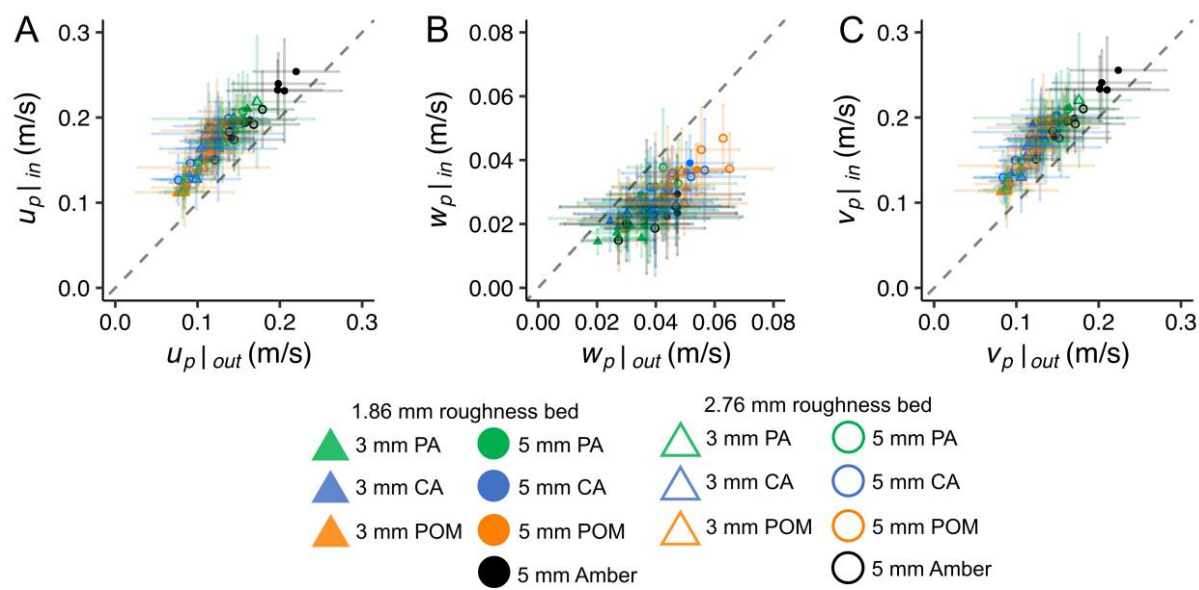

Figure 9. The average inwards and outwards collision velocities in the A) streamwise ($u_p|_{in}$, $u_p|_{out}$) and B) vertical direction ($w_p|_{in}$, $w_p|_{out}$) and C) the velocity magnitude ($v_p|_{in}$, $v_p|_{out}$).

A analysis similar to the one presented in Section 3.2 and 3.4 was undertaken to identify where observed $u_p|_{in}$, $u_p|_{out}$ and $w_p|_{in}$, $w_p|_{out}$ velocities are possibly drawn from either a Gaussian or a Gamma function using K-S tests, and whether those are different for MPs and natural sediments. For completeness, Figure S5 shows a compilation of the K-S test results for $u_p|_{in}$, $u_p|_{out}$ and $w_p|_{in}$, $w_p|_{out}$, respectively, showing where data was consistent with a Gaussian and/or Gamma function (p > 0.05). The results show that the distribution functions of the collision velocities were analogous to all plastic materials in 90 % of cases for $u_p|_{in}$ and $u_p|_{out}$,

---

and 66 % of cases for $w_p|_{in}$ and $w_p|_{out}$, with only three cases from all collision velocities (1 %) differing from all corresponding plastic experiments.

### 3.4.3 Restitution coefficient

Provided that part of the inward velocity is lost during the particle-bed impact, it is convenient to define a restitution coefficient ($e$), which can be also defined in terms of its streamwise ($e_x$) and vertical ($e_z$) components:

$$e_x = \frac{u_p|_{out}}{u_p|_{in}} \quad (5)$$

$$e_z = \frac{w_p|_{out}}{w_p|_{in}} \quad (6)$$

$$e = \frac{v_p|_{out}}{v_p|_{in}} \quad (7)$$

The restitution coefficient is directly connected to the loss of kinetic energy during impact (Schmeeckle et al. 2001; Zeeshan Ali and Dey 2019). Figure 10 shows the average restitution coefficients by flow condition in A) the streamwise direction $e_x$, B) vertical direction $e_z$ and C) magnitude $e$ for all particles, plotted against the Rouse number $P$. Values of $e$ are lower than 1 outlining that kinetic energy is lost during particle-bed collision, while values of $e_z$ are larger than 1, suggesting that kinetic energy is transferred from the streamwise to the vertical direction during collisions, as also indicted by results from the $\alpha_{in}$ and $\alpha_{out}$ collision angles (Figure 8A). Values of $e_x$, $e_z$ and $e$ for amber particles are similar to previous studies observations for natural sediment (Hu and Hui 1996a; Niño and García 1998).

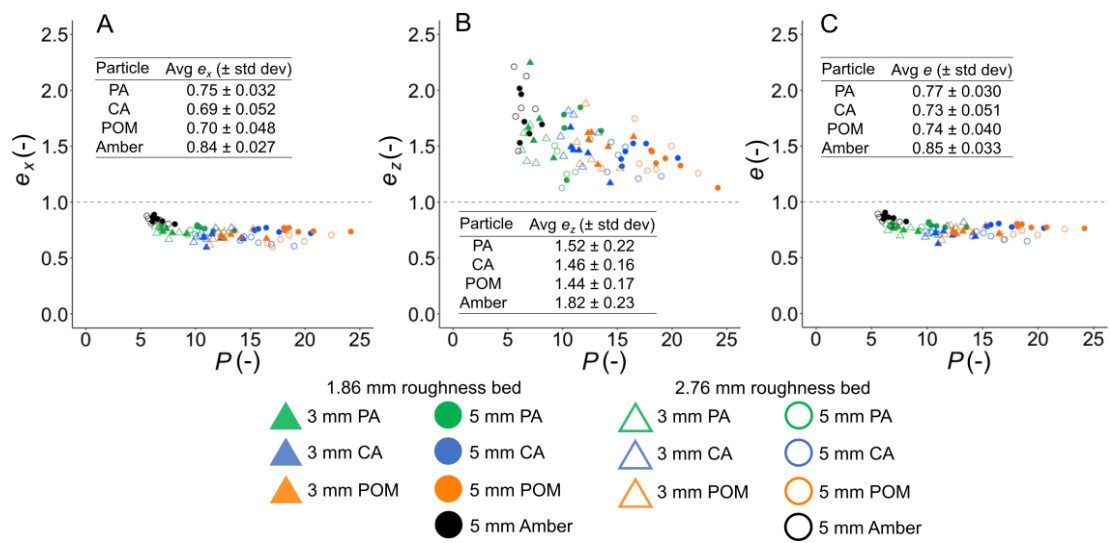

Figure 10. Restitution coefficients for PA, CA, POM and amber particles for the A) streamwise $e_x$, B) vertical $e_z$ direction and C) magnitude $e$.

Results suggest that values of $e_x$ and $e$ for MP particles may be independent of the Rouse number, while values of $e_z$ show a tendency to decrease with increasing Rouse number. It is observed that average values of $e_x$, $e_z$ and $e$ for PA, CA and POM particles are consistently lower than values for amber particles. An ANOVA statistical test revealed significant differences between values of $e_x$, $e_z$ and $e$ for the different materials (ANOVA, p <0.001) and a Tukey HSD post hoc confirmed the same findings (amber ~ PA, CA, POM, p <0.01). Overall, these results indicate that the MP particles lose more energy per successive saltation compared to amber particles at impact, which have higher Young's modulus than the MP particles (Table 2) and remain more elastic.

## 4    Conclusion

This study presents the results of novel laboratory experiments on bed load saltation dynamics using high-speed camera imaging and the detection of 11,035 individual saltation events to identify the similarities and differences between bed load transport dynamics of MPs and natural sediments. Our findings support the following conclusions:

- Saltation trajectory characteristics of MPs are analogous to natural sediments, as distribution functions for MPs were the same as natural sediment with only one amber experiment (1.4 % of cases) differing from all other plastic materials (Figure 6). In all cases, the Rouse parameter could explain saltation length, height and transport velocity equally for all materials tested.

- Inwards and outwards collision angles were well described by the Rouse number, with negligible material influence (Figure 8). Only nine experiments (9.3 % of cases) showed that distribution functions of impact angles for MPs differed from all natural sediment experiments.

- Differences in terms of restitution become negligible in overall transport dynamics as turbulence outweighs the kinetic energy loss during particle-bed collisions and keeps particle motion independent from impact (Figure 10).

To conclude, spherical MP particles behave similarly to spherical natural sediments in aquatic environments, within experimental uncertainty. Potential differences, due to particle shape will need to be tested separately in future studies since shape-deviations from sediments are significant and have yet to be investigated. The findings of this study are important because there is a well-recognised body of literature in bed load transport (van Rijn 1984; Garcia 2008;

Ancey 2020) that can be directly applied to the description of MP transport in rivers within the range of flow conditions herein tested.

**Acknowledgements**

This research has been funded by Erasmus+ and the UK Engineering and Physical Sciences Research Council (EPSRC) grant number EP/T517951/1. The authors are grateful to Matthew Moore for the support provided in data analyses.

---

---

---

---

---

---

---